\documentclass[10pt,aps,prb,psfrac,showpacs]{revtex4}
\usepackage{epsfig}
\usepackage{graphicx}
\usepackage{amsmath}
\usepackage{amsfonts,amssymb}

\def\to{\rightarrow}

\def\p{\partial}

\def\non{\nonumber }

\def\om{\omega}

\def\s{\sigma}
\def\med{\frac{1}{2}}
\newcommand{\beq}{\begin{equation}} 
\newcommand{\eeq}{\end{equation}} 
\newcommand{\beqa}{\begin{eqnarray}} 
\newcommand{\eeqa}{\end{eqnarray}}

\begin{document}

\title{Generalized effective hamiltonian for graphene under non-uniform strain}
\author{Juan L. Ma\~nes} 
\affiliation{Departamento de F\'isica de la Materia Condensada Universidad del Pa\'is Vasco, Apdo. 644, E-48080 Bilbao, Spain}
\author{Fernando de Juan}
\affiliation{Lawrence Berkeley National Laboratory,
1 Cyclotron Rd, Berkeley, CA 94720}
\affiliation{Department of Physics, University of California, Berkeley, CA 94720, USA}
\author{Mauricio Sturla} 
\affiliation{Instituto de Ciencia de Materiales de Madrid,\\
CSIC, Cantoblanco; 28049 Madrid, Spain.}
\author{Mar\'ia A. H. Vozmediano}
\affiliation{Instituto de Ciencia de Materiales de Madrid,\\
CSIC, Cantoblanco; 28049 Madrid, Spain.}
\date{\today}
\begin{abstract}
We use a symmetry approach to construct a systematic derivative expansion of the low energy
effective Hamiltonian modifying the continuum
Dirac description of graphene in the presence of non-uniform elastic
deformations. 
We extract  all experimentally relevant
terms and describe their physical significance.
Among them there is a  new gap-opening term that describes the Zeeman coupling
of the elastic pseudomagnetic field and the pseudospin. We 
determine the value of the couplings using a generalized tight binding model. 
\end{abstract}
%
%
%
%
 \maketitle

\section{Introduction}

The effects of lattice deformations on the electronic properties of graphene has been a topic of interest since the very early days due to the observation of ripples  in  suspended samples \cite{Metal07}. Later on, the subject acquired a new dimension after   the recognition of the extraordinary mechanical properties of the material \cite{LWetal09} and the capability of tailoring the samples to exploit the interplay of mechanical and electronic properties \cite{KZetal09,TKetal11}. The successful description of the influence of elastic deformations on the electronic excitations in terms of  ``elastic gauge fields" \cite{SA02,GHD08,VKG10} has been  extensively used in the proposals of strain engineering in real \cite{DWetal11} and synthetic samples \cite{GMetal12}. The interest in the effective low energy Hamiltonian of deformed graphene has been 
reactivated recently based on the apparent discrepancy between the lattice description --  tight binding (TB) approximation  and subsequent continuum limit  -- and an alternative geometric approach using the formalism of quantum field theory in curved spaces \cite{JCV07,JSV12}. There have also been recent claims of the emergence of new gauge fields in the standard TB approach originating from the deformation of the lattice vectors \cite{KPetal12,KPetal13,JMV13}. 

Given the rapid progress in this field one obvious question is, have we considered all possible effects of strain on the electronic properties of graphene or are we missing some? This is a crucial question, as  particular models and approximations tend to capture  specific  features of the physics  and, as a consequence, are likely to miss other aspects. We may answer this question by using group theory techniques to generate \textit{all} possible interactions respecting the symmetries of the system, and then try to find a model to estimate the values of their couplings.

The idea of constructing effective actions for physical systems based solely on symmetry considerations has a long tradition  both in quantum field theory (QFT) \cite{R97} and  condensed matter physics, and lies at the hearth of the Landau Fermi liquid theory of metals \cite{L57}. The Dirac description of the low energy electronic excitations of graphene in the continuum limit is rooted in the symmetries of the underlying honeycomb lattice, as has been known for a long time \cite{SW58}. The symmetry approach  has been applied to the  particular problem of strained graphene, for example, in refs. \onlinecite{M07,WZ10,FWetal11,Lin12,CGetal13}. A highly detailed symmetry construction has been used in ref.  \onlinecite{Basko08} to extract the 
low energy Hamiltonian affecting the Raman responses in graphene and, more recently, to explore the influence of the flexural modes on the spin--orbit coupling~\cite{FWetal11,OCetal12}. 

While many previous studies  have concentrated on uniform strains, important effects such as the emergence of pseudomagnetic fields~\cite{GKG10} and the new vector fields~\cite{JCV07,JSV12} responsible for pseudospin precession~\cite{JMV13} require the presence of non-uniform strain. Under non-uniform strain, new interaction terms arise which depend not just on the strain components, but also on their derivatives. In this work we  apply  standard symmetry based methods to construct a low energy effective hamiltonian  for  graphene in the presence of non-uniform elastic deformations. In order to accomplish this, we  set up a systematic expansion in derivatives of the strain and use group theory techniques to guarantee that  all the symmetry allowed terms  up to a given   order are included. Next we compute the coefficients of the most relevant terms --those which will affect the experiments -- within a generalized tight binding approximation, which sheds light  on the physical origin and significance of the various interactions in the effective hamiltonian. Those  terms which do not involve derivatives of the strain  have been already  discussed in the literature,  but  among the new interactions predicted by the symmetry approach there is one that opens a gap  and represents the Zeeman coupling between the elastic pseudomagnetic field and  pseudospin. We discuss the physical strength of this coupling within the generalized tight binding model and analyze some physical consequences.

The article is organized as follows. In Sec. \ref{secgeneral} we outline the properties of the graphene system  relevant to our symmetry analysis and set up a systematic expansion in the  derivatives of the strain tensor. Sec. \ref{hams} summarizes the  results of the symmetry analysis   and contains a description of all the possible terms  in the low energy Hamiltonian for deformed graphene with at most one derivative. The effects of including higher derivatives are explored in Subsec.~\ref{beyond}. In Sec. \ref{secTB} we introduce a generalized tight binding model which is used to compute the coefficients of the low energy Hamiltonian both  for in--plane strains and  out--of--plane distortions (\ref{secout}). We also consider the geometric terms due to frame effects (\ref{secframe}) and discuss some physical implications of the new gap opening term (\ref{secgap}). In Sec. \ref{secfinal} we summarize our work and consider possible extensions.
\section{Effective hamiltonian, derivative expansion and symmetries}
\label{secgeneral}

In this paper we consider a systematic expansion of the hamiltonian  in derivatives of the
 the electron field and the strain tensor~\footnote{ Such an expansion was already adopted in ref.  \onlinecite{JSV12} where the space dependent Fermi velocity was derived in the tight binding approximation.} 
 \beq\label{usual}
u_{ij}= \med(\p_i\xi_j+\p_j \xi_i+ \p_i h\p_j h)\;\;\;,\;\;\;i,j=x,y,
\eeq 
 where $\xi_i$ and $h$ are horizontal and vertical displacements respectively. This makes sense in the presence of elastic deformations, where each new derivative of the strain is expected to be suppressed by a factor of order $\mathcal O (a/\lambda)$, where $\lambda$ is the wavelength of the deformation and $a$ is  the lattice constant.  As we are interested in  a continuum low energy approximation where electrons behave like Dirac fermions,  we will restrict ourselves to terms that are at most linear in the electron momentum $k$, where $k$ is measured with respect to 
a Fermi point.  Moreover, we will assume that the system is  within the domain
of applicability of standard linear elasticity theory and consider only terms
linear in the strain tensor.  
 Thus the effective hamiltonian will be a  function of the electron fields $\psi$ and $\psi^\dagger$, the strain $u_{ij}$ and their derivatives. Each order in the derivative expansion will be characterized by $(n_q,n_k)$, where $n_q$ and $n_k$ count the order of the derivatives of the strain and electron fields respectively. Possible extensions of this approach to include nonlinear contributions and optical modes will be discussed in Sect.~\ref{secfinal}
\begin{table}[b]
\begin{tabular}{|c || c | c | c | c | c |}
\hline
$(n_q,n_k)$ &$(0,0)$&$(0,1)$&$(1,0)$&$(1,1)$&$(2,0)$\\
\hline \hline
$\openone$&1 & 1& 0& 0 & 2 \\
$\{\sigma_x,\s_y\}$ &1 & 2& 0& 0 & 3\\
$\sigma_z$ &0 & 0& 1& 2 & 0\\
\hline

\end{tabular}
\caption{ Number of independent  hermitian invariants linear in $u_{ij}$ at order  $(n_q,n_k)$ in the derivative expansion, containing the four $2\times 2$ hermitian matrices $\{\openone, \vec \s\}$. For each of these invariants  another one can be constructed through the substitution $u_{ij}\to \p_i h\p_j h$.}
\label{tn}
\end{table}

Any valid effective hamiltonian must respect all the symmetries of the system. In the case of graphene, these include the point group $D_{6h}$ of the honeycomb lattice~\footnote{In this paper we follow the  conventions and definitions in Ref.~\onlinecite{brad}.}. $D_{6h}$ consists of $24$ symmetry operations, and one of them is reflection by the horizontal plane $\s_h$. A first simplification is afforded by the fact that all the ingredients in the effective hamiltonian are invariant under reflection by  $\s_h$. More concretely, electron fields are combinations of $p_z$ orbitals which are  odd under $\s_h$, but only bilinears in the electron field are allowed in the hamiltonian and these are obviously even. Similarly, vertical atomic displacements $h$ are odd under $\s_h$, but only the combinations $(\p_i h) (\p_j h)$ enter the hamiltonian and these are even. As a consequence, we may ignore $\s_h$ as a symmetry and consider $C_{6v}$ instead of  $D_{6h}$. $C_{6v}$ has only $12$ elements, which include rotations by multiples of $\pi/3$ around the $OZ$ axis and reflections by six vertical planes.

 As is well known~\cite{NGetal09} the Fermi surface of the system  at half filling consists of six Dirac points located at the corners of the Brillouin zone in momentum space. Only two are non--equivalent, and can be chosen at opposite corners,   $K_2=-K_1$.  We will study the case where there are no interactions relating the two Fermi points and analyze each of them independently. Then the  low energy description of the electronic excitations around these points is governed by two Dirac Hamiltonians related by time reversal. This is the relevant situation for long wavelength elastic deformations, and in this case   the Dirac points are  protected against gap opening by smooth deformations respecting inversion and time reversal symmetry\cite{MGV07}.
Then symmetry allowed interactions around $K_1$ must be invariant only under the elements of $C_{6v}$ which leave $K_1$ invariant. This is known as the little point group~\cite{brad} of $K_1$, which is given by $C_{3v}$. As reviewed in  Appendix~\ref{apsym}, $C_{3v}$ is a subgroup of $C_{6v}$ with only $6$ elements: rotations by multiples of $2\pi/3$ around the $OZ$ axis and reflections by three vertical planes. Besides the little point group $C_{3v}$, $K_1$ is also invariant under the combined operation $C_2\theta$, where $C_2$ is a rotation by $\pi$ around the $OZ$ axis and $\theta$ is time reversal. Once a hamiltonian respecting $C_{3v}$ and $C_2\theta$ has been constructed around  $K_1$, time-reversal symmetry,   which takes $K_1$ into $K_2$, can be used to obtain the hamiltonian at $K_2$. This ensures that the total hamiltonian, which is the sum of the two hamiltonians around $K_1$ and $K_2$, respects all the symmetries of the system.

Once we know the set of symmetries to be respected by the interaction terms, the next step is to classify the relevant magnitudes according to their transformation properties. The result is shown in Table~\ref{t1} in Appendix~\ref{apsym}, where the relevant objects are assigned  irreducible representations of the little point group $C_{3v}$ and their  behaviour under $C_2\theta$ is indicated. Then one can use Eq.~\eqref{ninv} to determine the number of independent hermitian invariant  terms at each derivative order $(n_q,n_k)$ (see Table~\ref{tn}). This  crucial step  guarantees that all symmetry compatible interactions are taken into account. Then standard group theory techniques are used to construct all the symmetry allowed interactions.

\section{Symmetry-allowed terms in the Effective hamiltonian}
\label{hams}

\subsection{Effective Hamiltonian  to first derivative order}
%
\begin{table}[h]
\begin{tabular}{|c || c |c | c | c | }
\hline
$H_i$ & $(n_q, n_k)$ & Interaction term & Physical interpretation & $K_2$\\
\hline\hline
$H_1$ & (0,0) & $(u_{xx}+u_{yy})\openone$ &  Position-dependent electrostatic pseudopotential & $+$\\
  $H_2$ & (0,0) & $(u_{xx}-u_{yy})\s_x-2u_{xy} \s_y$& Dirac cone shift or U(1) pseudogauge field $(A_x , A_y)$ & $-$ \\
 \hline
$H_3$ & (0,1) & $\big[(u_{xx}-u_{yy})k_x-2u_{xy} k_y\big]\openone$ & Dirac cone tilt & $-$\\
$H_4$ & (0,1) & $(u_{xx}+u_{yy})( \s_x k_x+\s_y k_y)$ & Isotropic position-dependent Fermi velocity& $+$\\
$H_5$ & (0,1) & $u_{ij} \s_i k_j\;\; ;\;\; i,j=x,y$& Anisotropic position-dependent Fermi velocity & $+$\\
 \hline
$H_6$ & (1,0) & $\;\;\;\;\big[\p_y(u_{xx}-u_{yy})+2\p_x u_{xy} \big]\mathbf\s_z\;\;\;\;$ & Gap opening by non-uniform strain & $-$\\
\hline
\end{tabular}
\caption{ Effective low-energy electron-strain interactions allowed by symmetry.}
\label{tH}
\end{table}

The results of following the procedure outlined above and detailed in Appendix~\ref{apsym} may be  summarized 
in  an effective hamiltonian which contains all the symmetry allowed interactions  to first derivative order, i.e., for $n_q+n_k\le 1$. This is given by
\beq
H=H_0+\sum_{i=1}^6 a_i H_i+\sum_{i=1}^6 \tilde a_i \tilde H_i,
\label{genham}
\eeq 
where $H_0= v_F( \s_x k_x+\s_y k_y)$ and the terms  $H_i$ are given in Table \ref{tH}. The terms  $\tilde H_i$  are obtained from those  in Table~\ref{tH} through the substitution $u_{ij}\to \p_i h\p_j h$, where $h$ is the vertical displacement. The reason for the appearance  of the extra terms $\tilde H_i$ in the hamiltonian is that $\p_i h\p_j h$ transforms exactly like $u_{ij}$ under all the symmetries of the system. Thus for each invariant written in terms of $u_{ij}$ another one exists with $u_{ij}$ replaced by $\p_i h\p_j h$, and the coefficients of $H_i$ and $\tilde H_i$ have to be determined independently. This will be done in the next Section and, for the time being, we will refer to the more familiar $H_i$.  
At the end of this Section we will argue that the effective hamiltonian~\eqref{genham} probably captures all the experimentally relevant effects due to non-uniform strain.

Eq.\eqref{genham} gives the form of the first-quantized hamiltonian. The corresponding second-quantized hamiltonian operator  is  given by 
$\hat H\!=\! \int d^2 x\, \psi^\dagger H \psi$, where the symmetric convention for the derivatives acting on the electron fields is understood, i.e.,   
 \hbox{$\psi^\dagger k_i\psi\!\to\! -i/2(\psi^\dagger\overleftrightarrow{\p_i}\psi)\!\equiv\!-i/2(\psi^\dagger\p_i\psi -\p_i\psi^\dagger\psi)$}.
 For instance, $\hat H_5$ is given by 
\beq\label{h5}
\hat H_5=-\frac{i}{2}\int d^2 x\, u_{ij} ( \psi^\dagger\sigma_i\overleftrightarrow{\p_j}\psi), 
\eeq
where $\partial_j$ acts only on the electron fields. The  advantage of using the symmetric derivative convention is that any real expression in the electron momentum $k$, the strain (and its derivatives)  and  a  hermitian matrix will automatically give rise to a second-quantized hermitian operator. This simplifies the counting and construction of hermitian invariants. In this regard, it is important to realize that $(n_q,n_k)$ in Table~\ref{tH} gives the orders  of the derivatives when terms are written with the symmetric convention. See also comments around Eq.~\eqref{parts} below.

Table \ref{tH}  displays  all the hermitian,  symmetry-allowed terms of given orders  $(n_q,n_k)$ in the derivatives of the electron fields ($n_k$) and strain ($n_q$), as indicated in the second column. The remaining columns give   
 their physical interpretation and the relative sign of the  couplings at the two non-equivalent Dirac points. 
In what follows we will comment briefly on the physical significance of the various terms which, with the exception of $H_6$, have already been given~\footnote{Note that the term in Eq.~(45) of ref.~\onlinecite{WZ10} is proportional to the combination $2H_5\!-\!H_4$} in refs.~\onlinecite{WZ10,Lin12}:

\begin{itemize}
\item[$\bullet$] $H_1=(u_{xx}+u_{yy})\openone$: This term has the form of a scalar potential $\Phi\sim u_{xx}+u_{yy}$  and was already described in ref. \onlinecite{SA02}, where the coupling strength was  estimated  to be of order 4 eV for single layer
graphene. Its physical consequences have been explored in ref.  \onlinecite{LGK11}.

\item[$\bullet$] $H_2=(u_{xx}-u_{yy})\s_x-2u_{xy} \s_y$: Dirac cone shift in momentum space or $U(1)$ pseudogauge field $(A_x,A_y)\sim(u_{xx}-u_{yy},-2u_{xy})$. This term corresponds to  the well known elastic pseudogauge fields of the standard tight binding approach. It  has been used in the literature to propose all kinds of strain engineering and to fit experimental measurements of very intense  pseudomagnetic fields in corrugated graphene samples \cite{VKG10}. It has also been used to explain data in artificial graphene \cite{GMetal12}.

\item[$\bullet$] $H_3=\big[(u_{xx}-u_{yy})k_x-2u_{xy} k_y\big]\openone$: Dirac cone tilt. This term appears naturally in the description of the two dimensional organic superconductors \cite{KKetal07} which are described by anisotropic Dirac fermions. It also arises when applying uniaxial strain in the zigzag direction, a situation that has been discussed at length in the literature \cite{GFetal08,WGS08,MPetal09,CJS10}.

\item[$\bullet$] $H_4=(u_{xx}+u_{yy})( \s_x k_x+\s_y k_y)$: Isotropic position-dependent Fermi velocity \cite{JSV12}.

\item[$\bullet$] $H_5=u_{ij} \s_i k_j\;\; ;\;\; i,j=x,y$: Anisotropic  position-dependent Fermi velocity \cite{JSV12}. This term, together with $H_4$, was predicted to arise within the geometric modeling of graphene based on techniques of quantum field theory in curved space \cite{JCV07}. It was later obtained in a tight binding model by expanding the low energy hamiltonian to linear order in $q$ and $\xi$ \cite{JSV12,JMV13}. It comes together with a new vector field  $\Gamma_i$ that will be discussed  below. Since the Fermi velocity is the most important parameter in the graphene physics, this term affects all the experiments and will induce extra spatial anisotropies in strained  samples near the Dirac point \cite{GKV08,BP08,ZBetal09,Gazit09b,GTetal12,PAP11}.

\item[$\bullet$] $H_6=\big[\p_y(u_{xx}-u_{yy})+2\p_x u_{xy} \big]\mathbf\s_z$:  This is a very interesting term that suggests a  new gap-opening mechanism that has not been noticed previously. It can be seen as the  Zeeman coupling of pseudospin to the associated pseudomagnetic field $B_z= \p_x A_y-\p_y A_x$\cite{ZLetal13}. The magnitude of this new gap will be estimated in   Subsec. \ref{secgap} where we will explore its physical implications.

\item[$\bullet$] To first order in the derivative expansion we can also construct an  invariant involving the antisymmetric derivative of the displacement vector $\om=\p_x\xi_y-\p_y\xi_x$:
\beq\label{om01}
\om(k_x\s_y-k_y\s_x)=\om\epsilon_{ij}k_i\s_j,
\eeq
but, as shown in Ref.~\onlinecite{JMV13}, it can be eliminated by a local rotation of the pseudospinor $\psi\to\exp(-\frac{i}{2}\om \s_z)\,\psi$. Thus the effective hamiltonian~\eqref{genham} does not depend on $\om$.
\end{itemize}

Note that the new vector field $\Gamma_i$, which plays the role of the spin connection in the geometric formalism and goes together with the position-dependent Fermi velocity as discussed in \cite{JSV12,JMV13}, does not appear explicitly in  Table \ref{tH}. However,  if one uses integration by parts on \eqref{h5} to revert $\hat H_5$ to the more common asymmetric convention, the result is 
\beq\label{parts}
\hat H_5= -i\int d^2 x\,  \psi^\dagger\sigma_i[u_{ij} \partial_j+\frac{1}{2}\p_j u_{ij}]\psi,
\eeq
where we recognize the contribution $\frac{1}{2}\p_j u_{ij}$ to the vector field $\Gamma_i$. Similarly, the other piece of $\Gamma_i$  is obtained after integration by parts of $\hat H_4$. Thus, even though the symmetric derivative convention seems to eliminate  $\Gamma_i$ from the hamiltonian, this field will show up in the equations of motion, which involve precisely this integration by parts. This means that $\Gamma_i$ is a relevant field, giving rise to physical effects such as pseudospin precession~\cite{JMV13}.

We close this subsection with a comment on the last column in Table \ref{tH}. If we assume that Eq.~\eqref{genham} gives the hamiltonian around  the $K_1$ Dirac point, then the hamiltonian around $K_2$ is obtained by flipping the signs of the couplings $a_i$ and $\tilde a_i$ according to the last column. This assumes the use of the $(A_1, B_1, B_2, A_2)$ convention for the pseudospinors. In other words, whereas the first component of the pseudospinor around $K_1$ refers to the $A$-sublattice, the first component around $K_2$ refers to the $B$-sublattice. With this convention  the unperturbed hamiltonians $H_0$ are identical around the two Dirac points  and the three components of the pseudospin operator ---the three Pauli matrices, not just $\sigma_y$--- are odd under time reversal. 
See  Appendix~\ref{apsym} for a detailed explanation.

\subsection{Beyond first derivative order}
\label{beyond}
Eq.\eqref{genham} with Table \ref{tH} gives the most general effective hamiltonian containing at most one derivative of the electron field or the strain, i.e., for  $n_q+n_k\leq 1$. One can easily go to higher derivative orders. For instance, according to the last column in Table~\ref{tn}, at order $(2,0)$ there are 
two new invariants  proportional to the unit matrix and three containing $\s_x$ and $\s_y$. Comparing to $H_1$ and $H_2$ in Table~\ref{tH}, it is obvious that the new invariants represent second derivative corrections to the the electrostatic $\Phi\sim u_{xx}+u_{yy}$ and vector pseudopotentials $(A_x,A_y)\sim(u_{xx}-u_{yy},-2u_{xy})$. These corrections are  easily constructed using the techniques reviewed in Appendix~\ref{apsym} and are summarized in Table~\ref{t20}. 
\begin{table}[h]
\begin{tabular}{|c || c | c|}
\hline
$\delta \Phi$ & $\delta A_x$ & $\delta A_y$\\
\hline\hline
$(\p^2_x+\p^2_y)(u_{xx}+u_{yy})\sim \nabla^2 \Phi$ & $ \nabla^2 A_x $& $ \nabla^2 A_y$ \\
$\;\;(\p^2_x-\p^2_y)A_x-2\p_{xy} A_y\sim\p_{ij} u_{ij}\;\;$ &$(\p_x^2-\p_y^2) \,\Phi$ & $-2\p_{xy}\, \Phi$\\
 & $\;\;\;\;\p_x(\vec\nabla\cdot\vec A)\;\;\;\;$ &$\;\;\;\;\p_y(\vec\nabla\cdot\vec A)\;\;\;\;$\\
\hline
\end{tabular}
\caption{ Second derivative corrections to the scalar and vector pseudopotentials.}
\label{t20}
\end{table}
However,  these higher derivative corrections are likely to be masked by the  order zero contributions to the same physical phenomena, and their relevance to experiments may   be negligible.
This is actually the general trend. As shown in  Appendix~\ref{apsym}, invariance under the combined operation $C_2\theta$ implies that  terms proportional to the matrices $\{\openone, \s_x,\s_y\}$ must contain an even number of derivatives of the strain, whereas this number must be odd for terms proportional to $\s_z$. As a result,  corrections contain two more derivatives than the leading contribution and should be strongly suppressed, at least for reasonably smooth strains. 

This observation can be used to argue 
 that Eq.\eqref{genham} and Table \ref{tH} already give the most general effective hamiltonian describing the electronic properties of strained graphene, in the following sense:  any additional terms that we may construct will not give rise to qualitatively different physical phenomena, they will just provide higher order corrections in the expansion in derivatives of the strain, or in powers of the strain itself.    To show this, we first notice that the most general perturbation of the massless Dirac hamiltonian $H_0$  which is linear in the electron momentum $k$ must take the form
\beq\label{pert}
\delta H= \alpha_1 \openone +\alpha_x \sigma_x+\alpha_y \sigma_y+\alpha_z \sigma_z,
\eeq
where the functions $\alpha_i$ are at most linear in $k$, i.e., $\alpha_i=\alpha_i^{(0)}+\alpha_{ij}^{(1)}k_j$. Now, comparing with Table \ref{tH} we have $\alpha_1^{(0)}\sim u_{xx}+u_{yy}$, $\alpha_{1x}^{(1)}\sim u_{xx}-u_{yy}$, etc.
The only missing terms are those giving the $\mathcal O(k)$ contribution to  $\alpha_z$. According to Table~\ref{tn}, there are two terms of order $(1,1)$  that contribute to $\alpha_{zx}^{(1)}$ and $\alpha_{zy}^{(1)}$. These are easily constructed with the techniques reviewed in Appendix~\ref{apsym}, and are given by
\beq\label{sz}
\epsilon_{ij}k_i\p_ju_{kk}\s_z\;\;\;\; \mathrm{and} \;\;\;\; \epsilon_{ij}(k_l\p_i+k_i\p_l)u_{lj}\s_z.
\eeq
We note in passing that the first term can be written as $\vec \s \cdot(\vec k \times \vec \nabla\Phi)$ and has the form of a pseudospin-orbit coupling. Now, the unperturbed Dirac hamiltonian $H_0$ plus the two terms in Eq.~\eqref{sz}
give a hamiltonian of the form \hbox{$H\sim v_F\s_i k_i+\s_z b_i k_i$}, which squares to
\beq
 \mathcal{E}^2=v_F^2k^2+(b_i  k_i)^2
 \eeq
and  one can see that the effect of the new terms on the spectrum  is just a change in the Fermi velocity, which becomes anisotropic and position-dependent. In other words, they give higher order corrections to an effect already accounted for by  $H_4$ and $H_5$ at lower order. As these corrections would probably be very hard to measure experimentally, the effective hamiltonian given by Eq.\eqref{genham} is, in this  phenomenological sense,  complete.

We finish this Section with a comment on the local rotation $\om=\p_x\xi_y-\p_y\xi_x$. The results of using Eq.~\eqref{ninv} with $u_{ij}$ replaced by $\om$ are given in Table~\ref{tnw}, which shows  that only three invariant terms involving $\om$ can be constructed with $n_q+n_k\leq 2$. 
\begin{table}[h]
\begin{tabular}{|c || c | c | c | c | c |}
\hline
$(n_q,n_k)$ &$(0,0)$&$(0,1)$&$(1,0)$&$(1,1)$&$(2,0)$\\
\hline \hline
$\openone$&0 & 0& 0& 0 & 0 \\
$\{\sigma_x,\s_y\}$ &0 & 1& 0& 0 & 1\\
$\sigma_z$ &0 & 0& 0& 1 & 0\\
\hline
\end{tabular}
\caption{ Number of independent  hermitian invariants linear in $\om=\p_x\xi_y-\p_y\xi_x$ at order  $(n_q,n_k)$ in the derivative expansion containing the four $2\times 2$ hermitian matrices $\{\openone, \vec \s\}$.} 
\label{tnw}
\end{table}

The one of order $(0,1)$, which is given in Eq.~\eqref{om01}, has already been discussed. The two remaining  invariants are 
\beq
(k_x\p_x\,\om+k_y\p_y\,\om)\s_z \;\;\; \mathrm{and} \;\;\; 2(\p_{xy}\,\om)\s_x+(\p_x^2\,\om-\p_y^2\,\om)\s_y.
\eeq
The first one is of order $(1,1)$ and should be added to the two invariants in Eq.~\eqref{sz}. The last one, of order $(2,0)$, provides an additional correction to the pseudogauge fields. Our previous discussion suggests that the effects of these two terms will be hard to detect experimentally.  
\section{Generalized tight binding Hamiltonian}
\label{secTB}

In the last Section we have used symmetry arguments to construct the allowed  terms  in the low energy hamiltonian in the presence of strain, but symmetry alone can not fix the values of the coefficients. In this Section we will use a generalized tight binding (TB) model to estimate the numerical values of the couplings. See Appendix~\ref{apTB} for our conventions and details on the derivation of Eq.~\eqref{ham}.

Nearest neigbors (NN) interactions take place only between atoms belonging to different sublattices. As a consequence, the resulting hamiltonian contains only  off-diagonal contributions  and misses all the  terms proportional to the matrices $\openone$ and $\sigma_z$. In order to generalize the standard  NN-TB model, two new parameters are introduced: $-t_2$ is the hopping integral between next to nearest  neighbors (NNN) and V is the contribution of a nearest neighbor  potential to the on-site energy of an electron in a $p_z$-orbital. Recent calculations of the values of these parameters can be found in ref.  \onlinecite{KGF10}.

We consider first the simpler case of in-plane strain ($h(\vec r)\!=\!0$), and Fourier expand the atomic displacements $\vec \xi(\vec r)$ 
\beq
\vec \xi(\vec r)=\sum_{\vec q}\vec \xi(\vec q\,) e^{i \vec q \cdot \vec r}\;\;  \mathrm{with}\;\;\;\vec  \xi(-\vec q\,)=\vec \xi(\vec q\,)^*.
\eeq
The electron Bloch waves are given by 
\beq\label{Bloch}
\Phi_i(\vec k)=\frac{1}{\sqrt N}\sum_{\vec t} e^{i\vec k\cdot (\vec r_i+\vec t)} \varphi(\vec r-\vec r_i-\vec t),
\eeq
where $\varphi(\vec r)$ denotes a  $p_z$ atomic orbital,  $\vec r_i$  ($i=1,2$) are the positions of the two atoms in a reference  unit cell and the sum runs over the $N$ points $\vec t$ in the Bravais lattice.
Denoting by $\vec v_n$ and $\vec w_n$  the relative positions of nearest and next-to-nearest neighbors respectively, the matrix elements of the hamiltonian 
\beq\label{mat}
\delta H_{ij}(\vec q, \vec k\,)=\langle\Phi_i(\vec k+\med \vec q\,)|\delta H|\Phi_j(\vec k-\med \vec q\,)\rangle
\eeq
are given by  
\beqa\label{ham}
\delta H_{11}(\vec q, \vec k\,)&=&-2 i t'_2\sum_{n=1}^6 \vec \xi (\vec q\,)\cdot \hat w_n e^{i \vec w_n\cdot (\vec K_1+\vec k)}\sin\bigl(\frac{\vec q\cdot \vec w_n}{2}\bigr)+V'\sum_{n=1}^3 \vec \xi (\vec q\,)\cdot \hat v_n(e^{i \vec v_n\cdot \vec q}-1)\non\\
\delta H_{12}(\vec q, \vec k\,)&=&-2 i t'_1\sum_{n=1}^3 \vec \xi (\vec q\,)\cdot \hat v_n e^{i \vec v_n\cdot (\vec K_1+\vec k)}\sin\bigl(\frac{\vec q\cdot \vec v_n}{2}\bigr),
\eeqa
where $-t_1$ is the usual hopping integral between NN neighbors,  $\beta=\p( \log t_1) /\p( \log r)$, and the primes denote derivatives $\p/\p r$ that are always evaluated at the equilibrium positions\footnote{Note that we define $\beta$ without  the customary absolute value,  which implies $\beta<0$.}. $\delta H_{21}$ and  $\delta H_{22}$ are obtained from $\delta H_{12}$ and $\delta H_{11}$ respectively by making the replacement $\vec v_n\to -\vec v_n$. 
Note the symmetric split of the phonon momentum $q$ among the incoming and outgoing electrons in~\eqref{mat}, which in position space implies the symmetric derivative convention used in the last Section. 
Eq.~(\ref{ham}) is valid to all orders in the electron and phonon momenta $\vec k$ and $\vec q$, and the generalization to include any number of neighbors is obvious: one just has to add new terms, with $\vec v_n$, $\vec w_n$ replaced by the vectors corresponding to the new neighbors. See Appendix~\ref{apTB} for our conventions and details on the derivation of Eq.~\eqref{ham}.
\begin{table}[h]
\begin{tabular}{|c || c | }
\hline
 $a_1$ & $\;\;\;\;\frac{3\sqrt{3}}{2}t_2'a+\frac{3}{2}V' a\;\;\;\;$\\
$a_2$ & \;\;$\frac{\beta}{2a} v_F$\\
 $a_3$ & \!\!\!$-\frac{9\sqrt{3}}{4}t'_2a^2$ \\
$a_4$ & $\frac{\beta}{4} v_F$\\
$a_5$ & $\frac{\beta}{2} v_F$\\
$a_6$ & $\frac{3}{8}V'a^2$\\
\hline
\end{tabular}
\caption{Crystal frame couplings for $\{H_i\}$. $a$ is the distance between NN.}
\label{tTB}
\end{table}

Expanding~\eqref{ham} to the required powers of $\vec q$ and $\vec k$, and comparing with Table~\ref{tH} we get the  values for the in-plane electron-strain couplings listed in Table~\ref{tTB}. Note that these values do not include possible corrections originating from the deformation of the lattice vectors \cite{KPetal12,KPetal13,JMV13}. The reason is that we are using equilibrium atomic positions in our Bloch functions~\eqref{Bloch} or, in  the language of Ref.~\onlinecite{JMV13}, we are working in the ``crystal frame". The contributions from the deformation of the lattice vectors~\cite{KPetal12,KPetal13,sloan,barraza}, also known as ``lab frame effects"~\cite{JMV13},  will be incorporated in Subsec.~\ref{secframe}.

The values of $a_2$, $a_4$ and $a_5$ can be obtained within the usual NN-TB model and have been known for some time. 
As the terms $\tilde H_i$ vanish for $h\!=\! 0$, in order to compute the corresponding  coefficients $\tilde a_i$ we must consider off-plane strains.

\subsection{Tight binding computation for off--plane strains}

\label{secout}
\begin{figure}[b]
\begin{center}
\includegraphics[angle=0,width=0.4\linewidth]{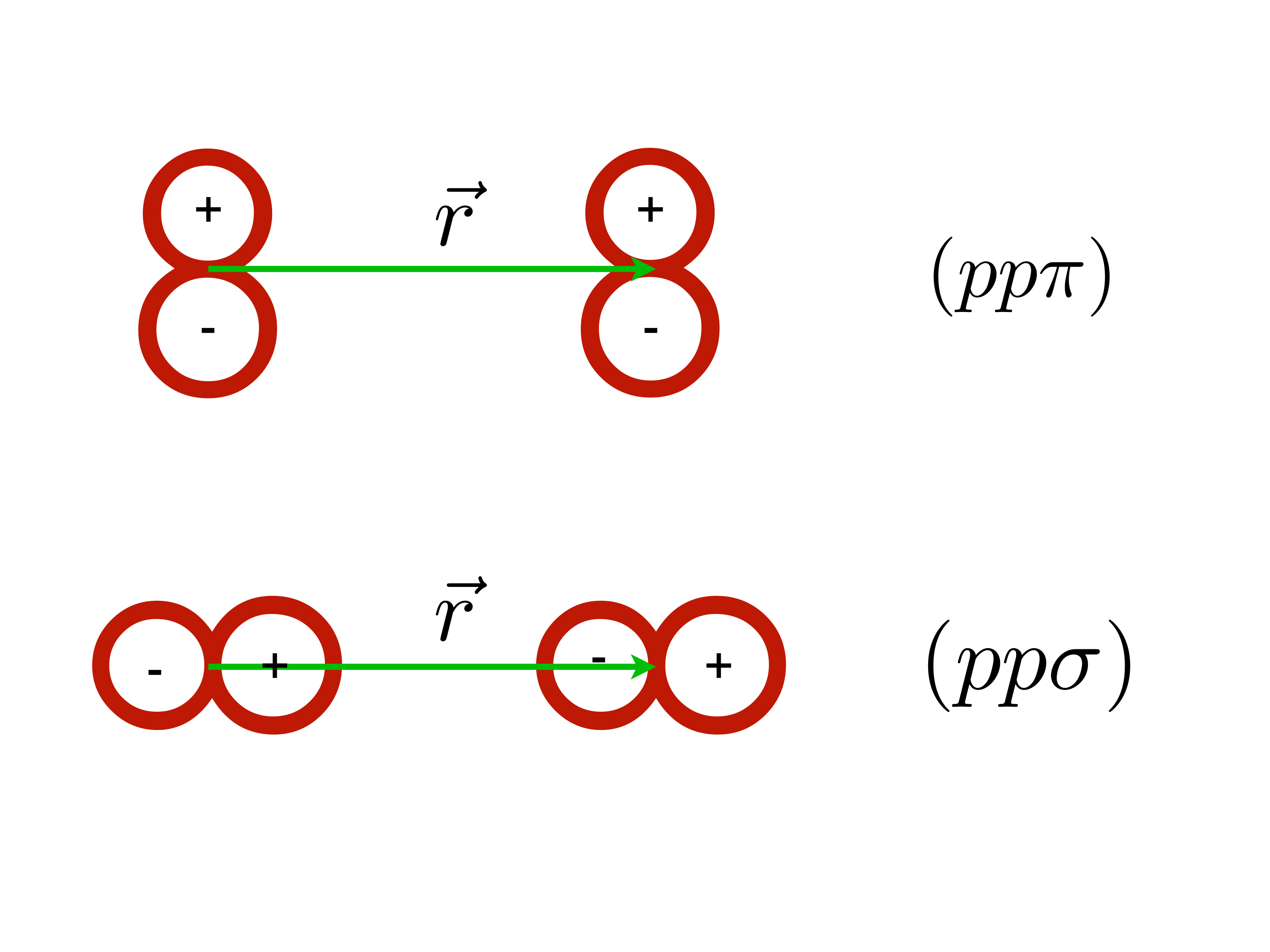}
\includegraphics[angle=0,width=0.4\linewidth]{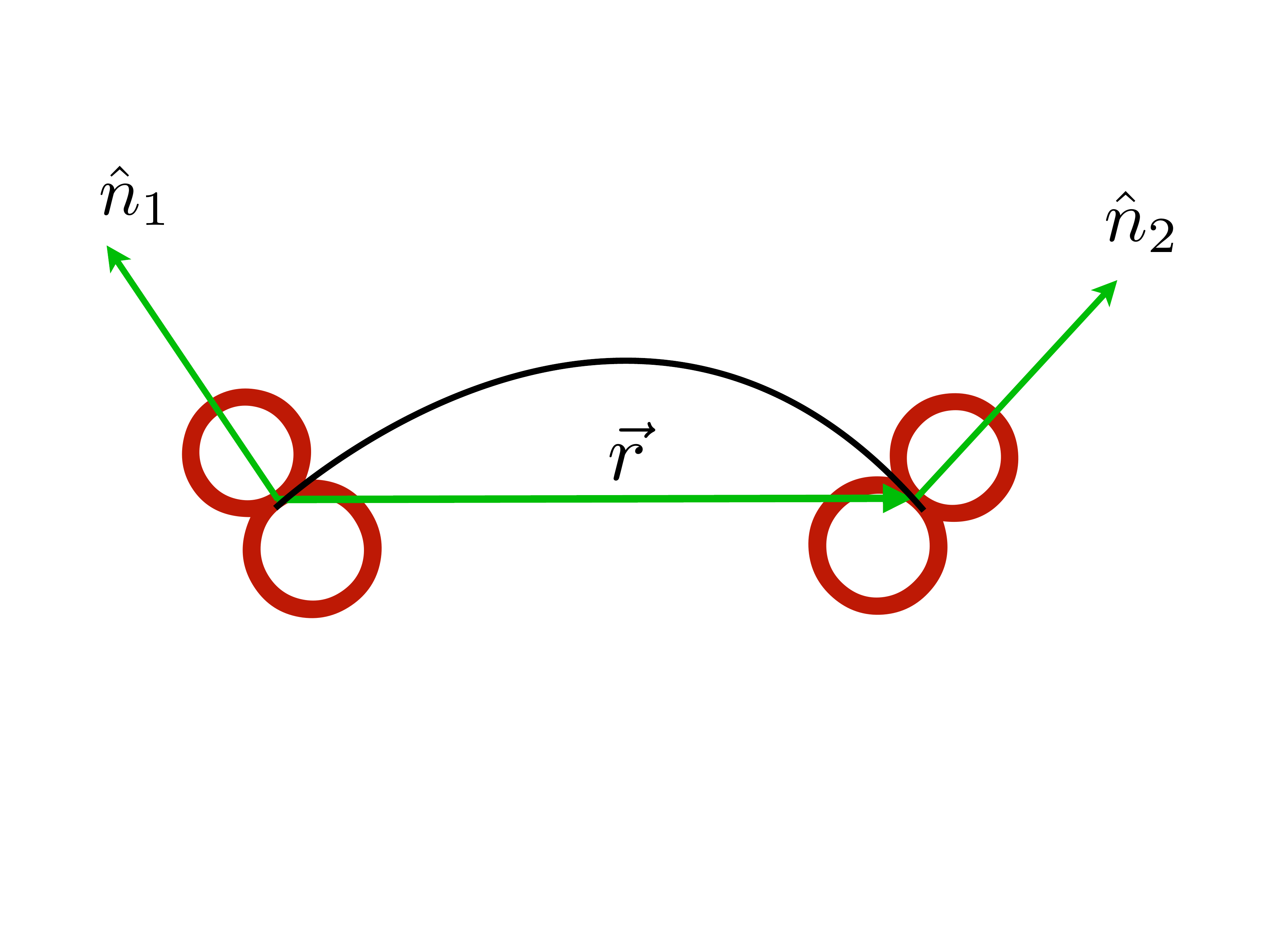}
\end{center}

\caption{Left: The two independent, $r$-dependent integrals $(pp\s)$ and $(pp\pi)$ for $p$-orbitals. For flat graphene, only the first one is relevant. Right: For graphene with curvature, $p$-orbitals are no longer parallel.}
\label{porb}
\end{figure}
The usual  assumption when dealing with non-planar strains is that off-plane atomic displacements $h(\vec r)$ enter the hamiltonian only through the combination ${u}_{ij}=(\p_i\xi_j+\p_j \xi_i+\p_i h\p_j h)/2$. 
The rationale  is that the distance between two nearby points is  given by 
$ds^2=(\delta_{ij}+2u_{ij})dx^i dx^j$
 where $dx^i$ is the difference between the equilibrium coordinates of the two points. 
However, this is be  justified if the  matrix elements between orbitals belonging to  different atoms depend only on the distance, which  is valid for  $s$-orbitals, but integrals involving $p_z$-orbitals are  non-isotropic.
To be concrete, whereas integrals involving two $s$-orbitals are parametrized by a single function of the distance, usually denoted $(ss\s)$, two independent functions  are required in the case of $p$-orbitals. These are denoted $(pp\s)$ and $(pp\pi)$, see Fig.~\ref{porb}. For flat graphene, only $(pp\pi)=-t_1$ is relevant, and in this respect $p_z$-orbitals behave just like $s$-orbitals.

However, in  presence of curvature the two $p_z$-orbitals are no longer parallel \cite{HGB06}. In terms of  $(pp\s)\equiv f_\s(r)$ and 
$(pp\pi)\equiv f_\pi(r)$, one can use linearity to show that the matrix element is then given by
\beq\label{ani}
\langle \Phi_1|H|\Phi_2\rangle=(\hat n_1\cdot \hat r) (\hat n_2\cdot \hat r) f_\s(r)+
[\hat n_1-(\hat n_1\cdot \hat r)\hat r]\cdot[\hat n_2-(\hat n_2\cdot \hat r)\hat r]f_\pi(r),
\eeq
where $\hat n_i$ are unit vectors parallel to the $p$-orbitals, which may be  assumed to be perpendicular to the surface. This has a rather involved dependence,  not only on $r$, but also on the angles. Thus, the assumption that the hamiltonian depends on $h$ only through Eq.~(\ref{usual}) is not valid in general for curved graphene.

On the other hand, in order to have curvature we need non-vanishing  second derivatives of $h$. This means that couplings involving only first derivatives of $h$ are independent of the  $(pp\pi)$ integrals and, as a result, their dependence on off-plane strains  is only through  $u_{ij}$. As only $H_6$ and $\tilde H_6$ involve second derivatives of $h$, we see that, with the usual approximations implicit in TB, $\tilde a_i\!=\! 0$ for $ i=1,\ldots, 5$.  Expanding eq. \eqref{ani} to the appropriate order it is easy to see that the first non-vanishing contribution is proportional to second derivatives of the strain and, as a consequence, the 
coefficient $\tilde a_6$ vanishes. 
This is true in the reference system of the perfect lattice (crystal frame).
Frame effects will be discussed in the next subsection.

\subsection{Lab frame effects}
\label{secframe}
Lab frame effects are the result of the change from crystal to laboratory coordinates, as  was discussed in detail in ref. \onlinecite{JMV13}. As a consequence, new terms independent of the TB couplings appear in the Hamiltonian. Crystal coordinates $\{ x\}$ are just the atomic equilibrium positions. If we  define the laboratory coordinates $\{ y_i\}$ as the horizontal projections of the out-of-equilibrium positions,  $y_i=x_i+\xi_i(x)$, then a change of variables in the continuum Hamiltonian   gives the result \cite{JMV13}:
\beq\label{main1}
\hat H_{Lab}=\hat H_{TB}+\hat H_{Geom},
\eeq
where $\hat H_{TB}$ is the hamiltonian in the crystal frame  and 
\beq
\hat H_{Geom}=
v_F\!\int d^2x\,\tilde u_{kl}(\psi^{\dagger}\s_k\overleftrightarrow{\p_l}\psi)\label{main2}
\eeq
with $\tilde u_{ij}= \med(\p_i\xi_j+\p_j \xi_i)$. Comparing with Table~\ref{tH}, we see that $\hat H_{Geom}$ is proportional to $\hat H_5$ with $u_{ij}$ replaced by $\tilde u_{ij}$. Thus both $a_5$ and $\tilde a_5$ are corrected to compensate for the absence of the non-linear piece in $\tilde u_{ij}$:
\beqa\label{labt}
\delta a_5= -\delta\tilde a_5= v_F.
\eeqa
 We have summarized our knowledge of the laboratory couplings in Table~\ref{tlab}. 
\vskip0.5cm
\begin{table}[h]
\begin{tabular}{||c | c || c | c || }
\hline
 $a_1$ & $\;\;\;\frac{3\sqrt{3}}{2}t_2'a+\frac{3}{2}V' a\;\;\;$ &  $\tilde a_1$ & $0$\\
$a_2$ & \;\;$\frac{\beta}{2a} v_F$& $\tilde a_3$ & $ 0$\\
 $a_3$ & \!\!\!$-\frac{9\sqrt{3}}{4}t'_2a^2$ & $\tilde a_2$ & $ 0$\\
$a_4$ & $\frac{\beta}{4}v_F$ & $\tilde a_4$ & $0$\\
$a_5$ & $(\frac{\beta}{2}+1)v_F$& $\tilde a_5$ & $\;\;-v_F\;\;$\\
$a_6$ & $\frac{3}{8}V'a^2$ & $\tilde a_6$ & 0\\
\hline
\end{tabular}
\caption{ Lab frame couplings for the effective hamiltonian.}
\label{tlab}
\end{table}

\subsection{Pseudo--Zeeman term}
\label{secgap}
As discussed in Sec.~\ref{hams},  $H_6$  is a new term which describes  the direct coupling of the $z$-component of pseudospin to the pseudomagnetic field $B$. This term plays the role of a mass in the Dirac fermion effective theory and opens a gap in the
spectrum. 
This mechanism is different from the various proposals of gap opening by strain in the literature, such as the the gap associated to the the Landau levels  \cite{LG10,GGetal10,LGK11,FWetal11}, superlattices \cite{PYetal08,Sny09,GL10,NB11}, or by merging of the Fermi points by  strain \cite{PCO09,CCC10}.
It is analogous to the one obtained by an on-site potential that is opposite in
the two sublattices (note that the required strain breaks inversion symmetry as well), but offers the
additional advantage of being tunable by the externally induced strain. This type of diagonal terms coming from strain have been recently discussed in ref.~\onlinecite{barraza} in an approach which uses directly 
the atomic displacements without reference to continuous elasticity theory.

The order of magnitude
of this gap may be estimated with the case of a ripple of moderate strain with height $h=5\mathrm{\AA}$ and
width $l = 25 \mathrm{\AA}$, which gives a pseudomagnetic field of
\beq
B\approx\frac{1}{l}\frac{h^2}{l^2}=0.0016\mathrm{\AA}^{-1},
\eeq
and an energy gap of the order of
\beq
E_{{\mathcal Zeeman}}=3/8 V' a^2 B\sim 7 meV,
\eeq
were we have taken the value $V'=6 eV/\mathrm{\AA}$ from ref.~\onlinecite{FWetal11}.
The presence of this new term has  several interesting implications. As it is known,
the orbital coupling of elastic pseudomagnetic fields comes with opposite signs in the two Fermi
points so that the combined effect of real and pseudomagnetic fields gives rise to valley separation
effects \cite{GKV08,GKG10,JLetal13,VAetal13}, and the same is expected for the Zeeman coupling. Indeed, in the presence of high magnetic fields, the zero-th Landau Level will be split by a controlled pseudo-Zeeman coupling and
induce valley polarization. In addition to providing a measurement of the coefficient $a_6$, this may
help to understand the origin of the observed interaction-induced splittings\cite{Goerbig12,PhysRevB.85.155439,YDetal12} --which can be of
similar magnitude at moderate field~\cite{BYM12}-- by studying the dependence of the gap with the pseudo--field,
and the competition with the real Zeeman coupling. As a related effect, the in-plane distortion
that generates this splitting may be induced spontaneously via a Peierls instability, by the same
mechanism as the out-of-plane distortion studied in ref. \onlinecite{PhysRevLett.98.016803}.

\section{Summary and discussion}
\label{secfinal}
 In this work we have used a symmetry approach to construct all possible terms affecting the low energy 
 properties of graphene in the presence of  non-uniform lattice deformations. We have limited our analysis to linear elasticity theory and assumed that the two Fermi points do not mix, which  is a sensible assumption for reasonably smooth strains.

 As we are primarily interested in the effects of non-uniform strain, we have  set up a derivative expansion of the effective hamiltonian and used   group theory techniques to obtain the number of independent couplings at each derivative order.  This procedure guarantees that no relevant interactions are  left out. Then we have  constructed the interactions in a completely model independent way. 
 
 After a careful analysis of the physical effects of the interactions and the properties of the derivative expansion, we have argued that the first order effective hamiltonian in Eq.~\eqref{genham} is ``phenomenologically complete", in the sense that any additional terms that we might construct would not give rise to qualitatively different physical phenomena, they would just provide higher order corrections. Under most experimental circumstances these corrections would be strongly suppressed and very hard to measure. 
 
In order to get an estimate of the values of the twelve coupling constants parametrizing the effective hamiltonian, we have considered a generalized tight binding model. This model incorporates, besides  first and second nearest neighbor hoppings, the contribution of a nearest neighbor potential to the on-site energy of an electron in a $p_z$-orbital. This contribution, which  is not often included in the tight binding hamiltonian, is necessary in order to account for the new gap-opening pseudo-Zeeman term coupling of pseudospin and pseudomagnetic field. This, and the fact that the pseudo-Zeeman term appears at first derivative order, while most tight binding computations are carried out for uniform strains, are the probable reasons why this term had gone unnoticed. This highlights the importance of the symmetry approach as a way to get all the allowed interactions in a model independent way.

In this paper we have neglected electron spin, but our analysis could be easily extended to accommodate it  along the lines of Ref.~\onlinecite{OCetal12}. Anharmonic effects are supposed to play an important role in the mechanical properties of graphene \cite{ZRetal10,COetal10,AGK12} although 
this assertion is yet to be confirmed by the experiments \cite{LH09}. On the other hand, their effects on the pseudomagnetic field has been  considered recently in Ref.~\onlinecite{RMP13} using a tight binding model. The techniques presented in this paper can be easily extended  to compute, in a model independent way,  all the allowed terms in an expansion in powers of the strain. 
Another possible extension is to include the effects of optical strains or frozen optical modes, which may affect the electronic properties of graphene on a substrate. Their effects on the pseudomagnetic fields at leading derivative order were considered in Ref.~\onlinecite{M07} and have been recently incorporated in an effective hamiltonian~\cite{Lin12}. Our methods could be used to explore their contributions at higer derivative orders.
 
\begin{acknowledgments}
We thank A. Cortijo, A. G. Grushin, H. Ochoa and E. da Silva for useful conversations. 
This research was supported by the Spanish MECD grants FIS2011-23713, PIB2010BZ-00512, FPA2009-10612, FPA2012-34456, the Spanish Consolider-Ingenio 2010 Programme CPAN (CSD2007- 00042),  by the Basque Government grant  IT559-10  and by NSF grant DRM-1005035. F. de J. acknowledges funding from the ``Programa Nacional de Movilidad de Recursos Humanos" (Spanish MECD).
\end{acknowledgments}

\appendix

\section{ The symmetry construction}
\label{apsym}

In this Appendix  we give a brief account of  the  group theory techniques used to construct the  effective hamiltonian.
As mentioned  in Section~\ref{secgeneral}, as long as we neglect interactions between the two inequivalent Dirac points we can restrict ourselves to the symmetries that leave $K_1$ invariant, i.e., to the little group of $K_1$.  The little point group $C_{3v}$  consists of six elements: the identity operation $E$, two $\pm2 \pi/3$ rotations $C_{3}^\pm$ around the $OZ$ axis and  three reflections $\sigma_{vi}$ by vertical planes. Transformation properties under $C_{3v}$ are classified by three irreducible representations (IRs):  $A_1$ and $A_2$ are one-dimensional, whereas $E$ is two-dimensional. Their character tables together with their products~\cite{brad} are given in  Table~\ref{CT}.

\begin{table}[h]
\begin{displaymath}
\begin{array}{c c}
\begin{array}{|c||c|c|c|}
\hline
C_{3v}& E & C_3^\pm& \sigma_{vi} \\
\hline
\hline 
A_1 & 1 &1&1 \\ \hline
 A_2 & 1 & 1 & -1\\\hline
E & 2 & - 1 & 0\\\hline
\end{array} \;\;\;\;\;\; & \;\;\;\;\;\;
\begin{array}{|c||c|c|c|}
\hline
C_{3v} & A_1 & A_2 & E \\\hline
\hline 
A_1 & A_1 & A_2 & E \\ \hline
 A_2 & A_2 & A_1 & E\\\hline
E & E & E & A_1 +A_2+E\\\hline
\end{array} 
\end{array} 
\end{displaymath}
\caption{Left: characters of the three irreducible representations of  $C_{3v}$. Right: decomposition of all possible product of two irreducible representations.} 
\label{CT}
\end{table}

Graphene is also invariant under time-reversal $\theta$, which takes the Dirac point $K_1$ into $K_2$, $\theta K_1\equiv K_2$. The same is accomplished by $C_2$, which  is $180$ degree rotation around the $OZ$ axis and belongs to the point group $C_{6v}$. Thus, the combined antiunitary operation $\theta C_2$ leaves $K_1$ invariant and imposes additional restrictions on the allowed interactions.

Table~\ref{t1} gives the transformation properties of all the ingredients used in the construction of the effective hamiltonian for  strained graphene. For the sake of completeness, we have  included the antisymmetric part of $\p_i\xi_j$, which represents a local rotation. Note that the transformation properties of  the Pauli matrices  follow from those of the electronic states. More concretely, the two components of the electron field $(\psi_1,\psi_2)$ transform according to the irreducible representation $E$. Then the set of four $2\times 2$ hermitian matrices belong to the reducible representation $E\times E$, which decomposes according to
\beq
E\times E=A_1(\openone)+A_2(\s_z)+E(\s_x,\s_y).
\eeq
\begin{table}[b]
\begin{tabular}{|c || c |c|  }
\hline
Magnitudes &\ IR of $C_{3v\;}$& $\ \theta C_2\;$ \\
\hline \hline
$\openone$, $u_{xx}+u_{yy}$& $A_1$  &$+$ \\
$\omega$ & $A_2$  & $+$ \\
$\sigma_z$ & $A_2$  &$-$\\
 $(u_{xx}-u_{yy}, -2u_{xy})$, $(k_x,k_y)$, $(\s_x,\s_y)$  & $E$  &$+$ \\
$(\p_x,\p_y)$  & $E$  & $-$ \\
\hline
\end{tabular}
\caption{ Transformation properties under the little group of $K_1$. The antisymmetric part of $\p_i\xi_j$ is given by the local rotation $\om=\p_x\xi_y-\p_y\xi_x$.}
\label{t1}
\end{table}
Group theory can now be used to obtain the number of independent terms in the effective hamiltonian at order $(n_q,n_k)$. The basic formula is~\cite{brad} 
\beq\label{ninv}
n=\frac{1}{N}\sum_g \chi_T (g),
\eeq
where $n$ is the number independent invariants, $N$ is the number of elements $g$ in the group and $\chi_T(g)$ is the character of $g$ in the representation $T$ associated to the interaction term. The character $\chi_T(g)$ is generally obtained as the product of the characters of the representations corresponding to the different components of the interaction term. As an example, assume that we want to know the number of independent terms of order $(n_q,n_k)=(0,1)$ containing the matrices $\s_x,\s_y$. This involves the quantities $u_{ij}$, $k_i$ and $\s_i$, which according to Table~\ref{t1} belong to the representations $A_1+E$, $E$ and $E$ respectively. Thus
\beq
\chi_T=(\chi_{A_1}+\chi_E)\times \chi_E\times \chi_E,
\eeq
which implies
\beq
\chi_T(E)=12 \;\;\; , \;\;\; \chi_T(C_3^\pm)=\chi_T(\s_{vi})=0.
\eeq
 Then Eq.~\eqref{ninv} gives $n\!=\!2$. 
 
 Note that according to Table~\ref{t1}, both $\s_z$ and the derivatives $\p_i$ acting on the strain are odd under $ \theta C_2$.
 Thus, terms proportional to the matrices $\{\openone, \s_x,\s_y\}$ must contain an even number of derivatives of the strain, whereas this number must be odd for terms proportional to $\s_z$. The results of using this method for $n_k+n_q\le 2$ are summarized in Table~\ref{tn} of Sect.~\ref{secgeneral}. The number of invariants involving $\om$ instead of $u_{ij}$ is given in Table~\ref{tnw}.

Invariant interactions, which by definition transform  like $(A_1, +)$, can  be  obtained by using  the following composition rules for the IRs of $C_{3v}$:
\beqa\label{comp2}
A_1(a)\times E(b_1,b_2)&=& E(a b_1,a b_2)\non\\
A_2(a)\times E(b_1,b_2)&=& E(a b_2,-a b_1)\non\\
E(a_1,a_2)\times E(b_1,b_2)&=& A_1(a_1 b_1+ a_2 b_2)+A_2(a_1 b_2-a_2 b_1)+E(a_1 b_1-a_2 b_2, -a_1 b_2-a_2 b_1).
\eeqa
For one-dimensional IRs, we  have $A_1(a)\times A_1(b)=A_1(ab)$, $A_2(a)\times A_2(b)=A_1(ab)$ and  $A_1(a)\times A_2(b)=A_2(ab)$. Several examples of the use of these realtions are given at the end of this Appendix.

Once an interaction term has been constructed around $K_1$, we can use the time reversal operation $\theta$ to construct the corresponding interaction around the other Dirac point $K_2$. Time reversal acts by complex conjugation, and its  action on the different objects is given in the l.h.s. of  Table~\ref{t2} for the the usual  $(A_1, B_1, A_2, B_2)$ sublattice  convention.

As an example, the Dirac hamiltonian $k_x \s_x+k_y \s_y $ at $K_2$ is given by
\beqa
\theta: k_x \s_x+k_y \s_y \to -k_x \s_x+k_y \s_y.
\eeqa
In order to compare interaction hamiltonians at the two Dirac points, we have to take into account that even $H_0$ differs by the sign of $k_x$. This fact, which makes a direct comparison  awkward, can be avoided by a change of basis at $K_2$. Conjugation of the Pauli matrices by $\s_y$ yields
\beqa
\s_y (\openone,\s_x,\s_y,\s_z)\s_y=(\openone,-\s_x,\s_y,-\s_z)
\eeqa
and now $H_0$ takes the same form at the two Dirac points
\beqa
\s_y(-k_x \s_x+k_y \s_y)\s_y= k_x \s_x+k_y \s_y.
\eeqa 
Conjugation by $\s_y$ changes the sublattice convention to  $(A_1, B_1, B_2, A_2)$. This is  summarized in the r.h.s. of  Table~\ref{t2}, which can be used to obtain  the hamiltonian at $K_2$ after conjugation by $\s_y$. Note that now all three Pauli matrices are odd under time reversal.

\begin{table}[h]
\begin{displaymath}
\begin{array}{c c}
\begin{array}{|c || c |}
\hline
\mathrm{Magnitudes} &\ \theta\; \\
\hline \hline
u_{ij},\om,  \p_i, \openone, \s_x, \s_z    & + \\
k_i, \s_y &  - \\
\hline
\end{array}\;\;\;\;\;\;\;\;\;\;\;\; & \;\;\;\;\;\;\;\;\;\;\;\;
\begin{array}{|c || c |}
\hline
\mathrm{Magnitudes} &\ \theta\; \\
\hline \hline
u_{ij}, \om, \p_i, \openone & + \\
k_i, \vec \s &  - \\
\hline
\end{array}
\end{array} 
\end{displaymath}
\caption{Transformation properties under  time reversal with the $(A_1, B_1, A_2, B_2)$ convention (left), and with the $(A_1, B_1, B_2, A_2)$ convention used in this paper (right).} 
\label{t2}
\end{table}

We finish this Appendix with a few examples:
\begin{itemize}
\item The fourth  line in Table~\ref{t1} together with the third line in Eq.~(\ref{comp2}) show that the Dirac hamiltonian \hbox{$H_0=k_x \s_x+k_y \s_y$} is invariant. Concretely,
\beqa
E(k_x,k_y)\times E(\s_x,\s_y)&=& A_1(k_x \s_x+k_y \s_y)+\ldots
\eeqa
\item The fourth line in Table~\ref{t1} together  the third line in Eq.~(\ref{comp2}) imply that 
$H_2=(u_{xx}-u_{yy})\s_x-2u_{xy} \s_y$ is invariant. Using Table~\ref{t2}  to obtain  $H_2$ at $K_2$  gives
\beqa
 (u_{xx}-u_{yy})\s_x-2u_{xy} \s_y \to -(u_{xx}-u_{yy})\s_x+2u_{xy} \s_y,
\eeqa
showing that  pseudogauge fields have opposite signs at the two Dirac points.
\item The fourth and fifth  lines  in Table~\ref{t1} together with  the third line in Eq.~(\ref{comp2}) show that~\hbox{$\p_y(u_{xx}-u_{yy})+2\p_x u_{xy}$} transforms according to  $(A_2,-)$. Concretely,
\beqa
E(\p_x,\p_y)\times E(u_{xx}-u_{yy},-2u_{xy})&=& A_2[\p_y(u_{xx}-u_{yy})+2\p_x u_{xy}]+\ldots
\eeqa
Then $A_2\times A_2=A_1$ and the third  line in Table~\ref{t1} imply that  \hbox{$H_6=\big[\p_y(u_{xx}-u_{yy})+2\p_x u_{xy} \big]\mathbf\s_z$}  is invariant (Zeeman coupling for pseudospin).

\end{itemize}

\section{ Generalized tight binding model}
\label{apTB}
In this Appendix we fix our conventions and give details on the tight binding model used to compute the coupling constants. We choose a coordinate system such that the vectors $\vec v_n$ to the three NN are given by
\beq
\vec v_1=a(0,1)\;\;\;,\;\;\; \vec v_2=-\frac{a}{2}(\sqrt{3},1) \;\;\;,\;\;\; \vec v_3=\frac{a}{2}(\sqrt{3},-1)
\eeq
where $a$ is the distance between NN. The vectors to the six NNN are given by
\beq
\vec w_1=-\vec w_4=-a(\sqrt{3},0)\;\;\;,\;\;\; \vec w_2=-\vec w_5=-\frac{a}{2}(\sqrt{3},3) \;\;\;,\;\;\; \vec w_3=-\vec w_6=\frac{a}{2}(\sqrt{3},-3)
\eeq
and the Fermi points are located at $\vec K_1=-\vec K_2=(\frac{4\pi}{3\sqrt 3 a},0)$ 

A general displacement
\beq
\vec \xi(\vec r)=\sum_{\vec q}\vec \xi(\vec q\,) e^{i \vec q \cdot \vec r}\;\;  \mathrm{with}\;\;\;\vec  \xi(-\vec q\,)=\vec \xi(\vec q\,)^*
\eeq
induces a change in the  vectors that go from an atom at position $\vec t+\vec r_1$ to its nearest neighbors   
\beq
\delta \vec v_n=\sum_{\vec q}\vec \xi(\vec q\,) e^{i \vec q \cdot (\vec t+\vec r_1)}\left[ e^{i\vec q\cdot \vec v_n}-1\right],
\eeq
with a similar expression for $\delta \vec w_n$. To linear order in $\vec \xi_i$, this induces a change in the NN hopping integral 
\beq\label{dt1}
\delta t_1(\vec t)= \vec \nabla t_1\!\cdot\delta \vec v_n=t_1'\,\sum_{\vec q}(\vec\xi(\vec q\,)\cdot\hat v_n) e^{i \vec q \cdot (\vec t+\vec r_1)}\left[ e^{i\vec q\cdot \vec v_n}-1\right]
\eeq
with analogous expressions for $\delta t_2$ and $\delta V$. Then, substituting~\eqref{dt1} into 
\beq
\delta H_{12}(\vec q, \vec k\,)=\langle\Phi_1(\vec k+\med \vec q\,)|\delta H|\Phi_2(\vec k-\med \vec q\,)\rangle=-\frac{1}{N}\sum_{\vec t,n} \delta  t_1(\vec t)e^{-i \vec q \cdot (\vec t+\vec r_1)} e^{i(\vec k-\frac{\vec q}{2})\cdot \vec v_n}
\eeq
and doing the sum over the Bravais lattice vectors $\{\vec t\}$ yields
\beq
\delta H_{12}(\vec q, \vec k\,)=-2 i t'_1\sum_{n=1}^3 \vec \xi (\vec q\,)\cdot \hat v_n e^{i \vec v_n\cdot (\vec K_1+\vec k)}\sin\bigl(\frac{\vec q\cdot \vec v_n}{2}\bigr).
\eeq
The same method is used to obtain the other matrix elements.
\bibliography{Symmetry3}

\end{document}